**Multiple first-order metamagnetic transitions and quantum oscillations in ultra-pure $Sr_3Ru_2O_7$**

R. S. Perry[1,2], K. Kitagawa[2], S. A. Grigera[3], R. A. Borzi[3], A. P. Mackenzie[3], K. Ishida[2] and Y. Maeno[1,2]

[1] *Kyoto University International Innovation Center, Kyoto 606-8501, Japan.*
[2] *Department of Physics, Kyoto University, Kyoto 606-8502, Japan.*
[3] *School of Physics and Astronomy, University of St. Andrews, St. Andrews KY16 9SS, UK.*

We present measurements on ultra clean single crystals of the bilayered ruthenate metal $Sr_3Ru_2O_7$, which has a magnetic-field-tuned quantum critical point. Quantum oscillations of differing frequencies can be seen in the resistivity both below and above its metamagnetic transition. This frequency shift corresponds to a small change in the Fermi surface volume that is qualitatively consistent with the small moment change in the magnetisation across the metamagnetic transition. Very near the metamagnetic field, unusual behaviour is seen. There is a strong enhancement of the resistivity in a narrow field window, with a minimum in the resistivity as a function of temperature below 1 K that becomes more pronounced as the disorder level decreases. The region of anomalous behaviour is bounded at low temperatures by two first-order phase transitions. The implications of the results are discussed.

Magnetic field tuning of quantum critical points in itinerant electron systems has generated considerable interest in recent years. The effect of field-tuned fluctuations has been studied in a number of systems including $URu_2Si_2$ [1,2], $YbRh_2Si_2$ [3], $CeRu_2Si_2$ [4,5], $UPt_3$ [6] and Fe/Cr multilayers [7]. In most studied systems, the quantum critical point results from the 'standard method' of using the field to depress the transition temperature of a second order phase transition. A condition for this approach is that the symmetry of the low-field phase be different from that of a field-polarised paramagnet. Recently, it has also been noted that quantum criticality can result in situations such as itinerant electron metamagnets in which both high- and low-field phases have the same symmetry [8]. In this case, the quantum critical point is the end point of a line of symmetry-conserving first order transitions [8,9].

The material that first stimulated the discussion of the quantum critical end-point is the subject of this paper, the bilayer ruthenate $Sr_3Ru_2O_7$. It is the $n$=2 member of the Ruddlesden-Popper series of layered perovskites $Sr_{n+1}Ru_nO_{3n+1}$, whose properties are dominated by layers of ruthenium oxide octahedra. The 4 electrons in the ruthenium 4d $t_{2g}$ orbitals hybridise with the oxygen 2p to create metallic bands that are narrow but still sufficiently broad to avoid insulating transitions in the majority of the materials. High purity was first reached in the $n$=1 member of the series, $Sr_2RuO_4$, enabling the observation of an unconventional superconducting state [10] in which there is good evidence for spin triplet pairing [11]. As the purity of $Sr_3Ru_2O_7$ was improved, it was realised that its ground state was paramagnetic [12] rather than ferromagnetic as had initially been reported [13]. High-field studies then revealed the phenomenon of metamagnetism, i.e. a sudden superlinear rise in magnetisation as a function of applied magnetic field [14]. Analysis of transport and specific heat gave evidence for quantum critical fluctuations, except for anomalous behaviour in the temperature dependent resistivity very near to the critical field $H_c$ for magnetic fields applied parallel to the crystallographic $c$ axis [8,14]. We concentrate on this orientation of the field throughout the present paper.

The anomaly seen near $H_c$ was a small change to the temperature dependent part of the resistivity. Very near to quantum critical points, the resistivity in itinerant systems often varies as $T^{\alpha}$, where the exponent $\alpha$ crosses over from some value less than 2 to become equal to 2 as Fermi liquid properties are recovered below some characteristic temperature $T^*$ [15]. In $Sr_3Ru_2O_7$, a value of $\alpha$ *higher* than 2 was seen below 0.8 K for fields $\mu_o H_c \pm 0.05$ T [8] (see also inset to Fig. 1). This observation was surprising because it could not be a simple consequence of critical fluctuations, but the experimental signal was very small. In common with many other metamagnetic systems, the residual resistivity of $Sr_3Ru_2O_7$ shows a pronounced peak at $H_c$, but the anomalous temperature dependent part constituted only about 3% of the total resistivity at 0.1 K. Many questions therefore remain open. Do the observations represent a violation of Matthiessen's rule in the presence of a strongly peaking resistivity, or are they the first hint of more profoundly unusual behaviour near $H_c$?

A crucial issue, of relevance far beyond $Sr_3Ru_2O_7$, is the role of disorder near metallic quantum critical points [15,16]. In a naïve approach, the most obvious way to study this would be via the deliberate introduction of disorder, since the existing crystals of $Sr_3Ru_2O_7$ were clean by the standards of the field, with values of the residual resistivity of less than 3 μΩcm. Ruthenate physics is, however, known to be extremely sensitive to disorder [17], so we have taken the opposite approach of striving to *decrease* the levels of disorder still further. As we will show, this has resulted in remarkable qualitative changes in the behaviour near the metamagnetic field.

The $Sr_3Ru_2O_7$ crystals used for this work were grown using a floating zone technique at Kyoto and measurements were made on several crystals from different growth batches at both Kyoto and St Andrews Universities. In Fig. 1 we show the effect on the magneto-resistivity of decreasing the residual resistivity in zero field ($\rho_{res}$) from 2.8 μΩcm to 0.4 μΩcm. For this direction of the applied field (parallel to the crystallographic *c* axis), a single peak is seen in the magnetoresistance for $\rho_{res} = 2.8$ μΩcm. As the sample purity increases, the same basic peak remains as the dominant feature, but more structure develops. In particular, there is a pronounced dip at 7.5 T, and the central peak is seen to have very steep sides at 7.8 T and 8.1 T.

The profound effect that decreasing the disorder levels is having is emphasised by the data shown in the inset to Fig. 1, which shows the result of sweeping the temperature at fixed fields chosen to be at the resistive maximum for each purity level. In the inset, the top trace shows the unusual $T^{-3}$ dependence previously reported for $\rho_{\mathrm{res}}$ = 2.8 μΩcm [8]. As the samples become purer, the temperature dependence of the resistivity at metamagnetic transition actually changes to a negative gradient at low temperatures, with a minimum that increases systematically in temperature as the purity increases. The anomalous flattening of $\rho$ can now be accounted for as the onset of a very weak minimum that has been depressed by the disorder to below 50 mK. A set of temperature sweeps at fixed field for one of our purest samples is presented in Fig. 2, to emphasise that the region of the $(H,T)$ plane in which the strange behaviour is seen is extremely small. Above 1.2 K, $\rho$ depends only weakly on field. Below that temperature, however, the behaviour at the peak differs remarkably from that on either side. Below 7.8 T and above 8.1 T, the resistivity has a positive gradient ($\partial\rho/\partial T > 0$) all the way down to 50 mK, crossing over to $T^2$ behaviour similar to that previously reported for the less pure samples [8]. In between these fields, however, a clear negative gradient is seen at low temperatures ($\partial\rho/\partial T < 0$). In our purest samples, the anomalous part is now very large (more than 30% of the total $\rho$ at 0.1K).

The almost step-like changes in the resistivity shown for the purest sample in Fig. 1 are reminiscent of the features expected at a first-order phase transition. To investigate them in more depth, we performed a study of the a.c. magnetic susceptibility $\chi(\omega)$. As discussed in ref. [18], the a.c. susceptibility is a sensitive probe for hysteretic dissipation through the appearance of an imaginary part $\chi''(\omega)$. The study of the previous generation of samples with $\rho_{\mathrm{res}} > 2.5$ μΩcm had shown that for an applied field in the crystallographic *ab*-plane, a single first-order transition line terminated in a critical point at approximately 1.25 K. As the field was rotated towards the *c*-axis, the end-point temperature was smoothly depressed to below 50 mK at around 10° from the *c* axis. No peak in $\chi''$ was observable for the field parallel to the *c* axis where the $T^3$ scattering rate was also observed at the metamagnetic field.

Like $\rho$, $\chi$ in the purest samples differs markedly from that in the previous generation of crystals near the metamagnetic field [17]. The resistivity and susceptibility of samples with $\rho_{\mathrm{res}} < 0.5$ μΩcm are shown in Fig. 3. There are three general correlations between $\rho$ and $\chi$. The first two correlations are the two steep features in $\rho$ coinciding with peaks in both $\chi'$ and $\chi''$. The peaks in $\chi''$ are particularly significant, as they indicate the presence of dissipation that, as argued in ref. [18], is most easily interpreted in terms of hysteresis at a first-order phase boundary. The data shown were taken at a temperature of 20 mK, but a series of sweeps at constant temperature in steps of 100 mK established that the dissipative peaks at 7.8 T and 8.1 T are observable up to ~800 mK and ~400 mK respectively. The third apparent correlation between the susceptibility and the resistivity is between a peak in $\chi'$ and onset of a *dip* in resistivity at ~7.5 T.

A summary of the findings presented in Figs. 1-3 is therefore as follows. Remarkably, *decreasing* the levels of disorder leads to an *increase* of the extra low temperature resistivity near the metamagnetic field, and to the resolution of several pronounced features in $\rho$ instead of the single peak seen in samples with $\rho_{res}$ = 2.8 μΩcm. In $\chi$, three peaks are seen at low *T*, two of which signal the crossing of first-order phase boundaries.

At first sight, these observations question the use of pictures based on itinerant electron magnetism [19] as the correct starting point for considering the physics of $Sr_3Ru_2O_7$. The basic idea of such theories is that the moment change at the metamagnetic field is due to a field-induced non-linear exchange split of the Fermi surface. The consequence of this should be a change of Fermi surface topography between the low-field and high-field states due to differences in the spin-up and spin-down Fermi volumes, but without a large overall change in *total* Fermi volume. The dramatically improved purity of the new samples allows us to address this issue via the first observation of quantum oscillations reported for $Sr_3Ru_2O_7$. In Fig. 4 we show tiny Shubnikhov-de Haas oscillations (less than 0.2 % of the total resistivity) from the field regions above and below the metamagnetic transition. The temperature is 0.1 K with the magnetic field parallel to the *c* axis. Frequencies corresponding to large Fermi surface sheets are observed in both cases, and all observed frequencies undergo changes of ± 10 – 30 % across the metamagnetic transition. The Fermi surface of $Sr_3Ru_2O_7$ is likely to be complicated [20], so a full analysis of the quantum oscillations is a separate project [21]. The results presented here already give strong support for the basic starting point of itinerant magnetism [8,9,15,18,19], because the size of the relative Fermi volume changes seems consistent with the low moment change at the metamagnetic transition of approximately 0.25 $\mu_B$ / Ru [22].

Although the underlying picture of physics based on itinerant electron metamagnetism is appropriate for $Sr_3Ru_2O_7$, the detailed behaviour near $H_c$ is not consistent with the simplest theories that are based only on a quantum critical end-point. The key experimental findings are that the enhancement of $\rho$ in the metamagnetic region grows as the disorder scattering decreases, and that the narrow field range over which the residual resistivity is enhanced seems to be bounded at either side by first-order phase transitions. Neither of these is predicted by, for example, ref. [15]. Enhanced scattering at first order phase transitions in metallic systems has been seen in other situations, notably in low carrier density materials such as quantum Hall ferromagnets [23]. In these systems, there are multiple peaks in the resistivity as a function of magnetic field corresponding to the crossing of Landau levels; in $Sr_3Ru_2O_7$ we observe only one peak. Also in contrast with those observations, the resistivity in this case remains high in the field range *between* the two transitions. Qualitatively, the magnetoresistance has similarity to that of graphite as it undergoes a field-dependent charge density wave transformation over a finite range of applied fields [24]. There, however, the change of resistivity is due to a large reduction in the Fermi volume in the density wave state, and there seems to be no reason to expect such a state in $Sr_3Ru_2O_7$. We do not resolve quantum oscillations in the region 7.8 T to 8.1 T, but in the high field state the Fermi volume seems to have made, at best, a small change from that at low fields. We also note that the resistivity plateau could be due to a change in the Fermi velocity close to the critical field.

We have currently propose no definite interpretation for the behaviour near $H_c$, but are interested in further investigation of several possibilities. The first is that the data signal the entry of the system into some new ordered state that it adopts to avoid the consequences of the divergent fluctuations near the critical point. In this sense, the physics of the critical point dictates the behaviour of the system, but theories based purely on the consequences of the associated fluctuations would be valid only outside the field range very near the metamagnetic transition. This reasoning (and the evidence for it) has been discussed in relation to $Sr_3Ru_2O_7$ [8] and $URu_2Si_2$ [1,2].

Another possibility is that the first-order physics now being seen in the best samples indicates the presence of a Griffiths-like phase [25] in which the high resistivity is dominated throughout the peak by the presence of magnetic domains. We see two distinct peaks in $\chi''$, but it may be possible to account for that in such a picture. Also, we cannot rule out the presence of a small dissipative signal or even static domains in the intermediate region. We plan to test for this in future with experiments involving field modulation of the resistivity.

In summary, we have reported a series of measurements on extremely pure single crystals of $Sr_3Ru_2O_7$ with residual resistivity as low as 0.4 μΩcm. The extreme purity of the samples has enabled the observation of quantum oscillations both above and below the metamagnetic field $H_c$. More surprisingly, it has also highlighted intriguing qualitative changes to the behaviour very near $H_c$. In the best samples, a well-defined region bounded by first-order phase transitions can be seen at low temperatures. Within this region, an upturn in the temperature dependence of the resistivity is observed below 1 K. We believe that this provides a striking example of the extreme sensitivity of correlated electron phenomena to disorder in itinerant systems. It is sobering to realise that the previous generation of $Sr_3Ru_2O_7$ crystals, in which the physics reported here was almost entirely unobservable, were already pure in comparison to many of the materials currently being used to study quantum criticality and its consequences.

We are pleased to acknowledge useful discussions with H. Fukazawa, A. G. Green, S. R. Julian, N. Kikugawa, G. G. Lonzarich, T. Sakakibara, L. Taillefer, T. Tayama, H. Yaguchi, S. Nakatsuji, Kosaku Yamada and S. Fujimoto. We also thank the Japan Society for the Promotion of Science (RSP is a fellow of JSPS), the Leverhulme Trust, the UK Engineering and Physical Science Research Council and the Royal Society for valuable financial support. This work has also been supported by Grants-in-Aid for Scientific Research from the Japan Society for Promotion of Science and from the Ministry of Education, Culture, Sports, Science, and Technology (MEXT).

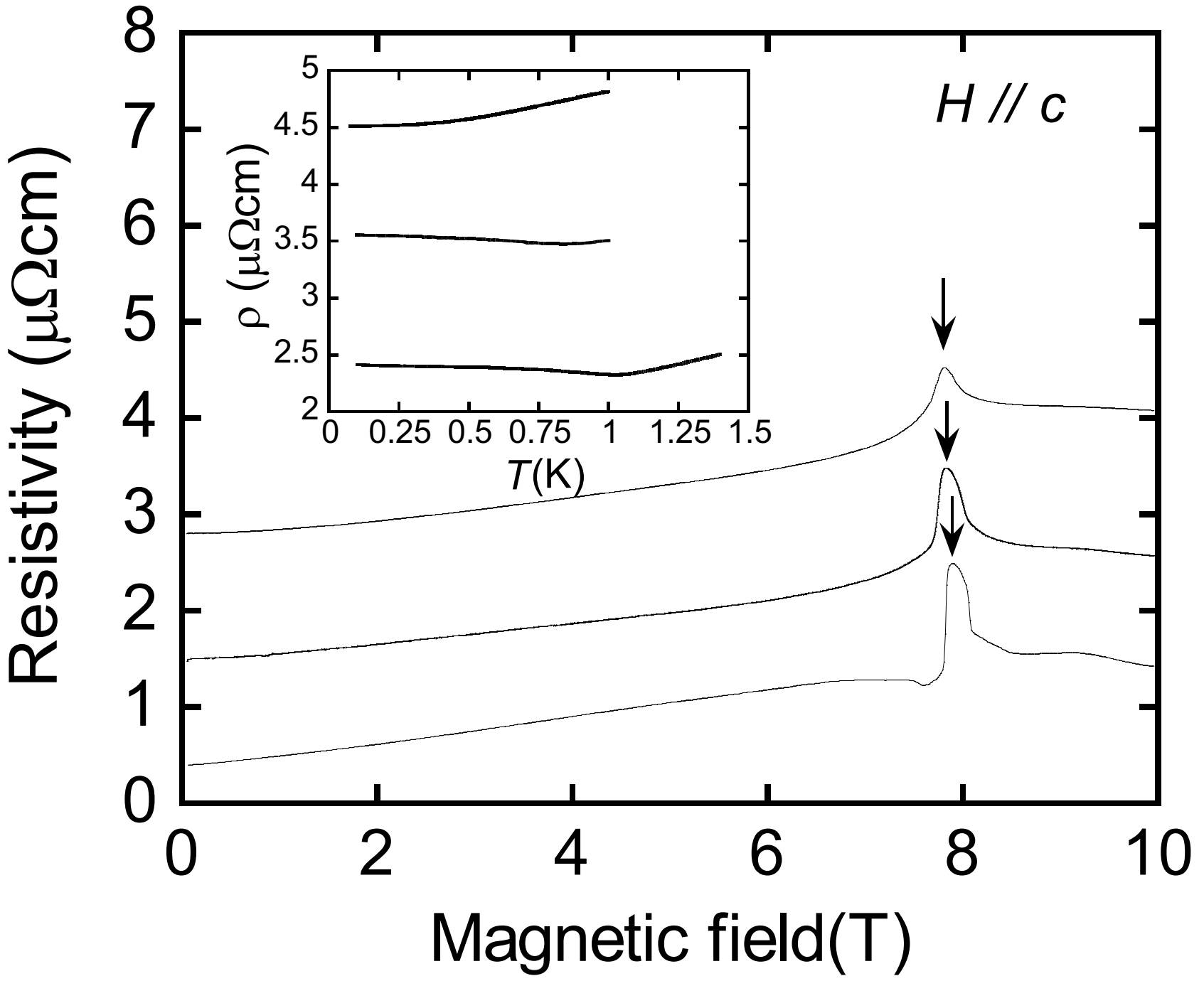

Figure 1. The resistivity ($\rho$) versus magnetic field in single crystals of $Sr_3Ru_2O_7$ for different values of residual resistivity $\rho_{res}$. The temperature of each curve is between 0.06 K and 0.1 K and the magnetic field is perpendicular to the $RuO_2$ planes. As the disorder level decreases, the resistive peak becomes higher, and very sharply bounded. Inset: Temperature sweeps at the fields indicated by the arrows in each case. Only the top curve (from ref. [8]) maintains a positive value of $\partial\rho / \partial T$. As the purity is increased, a well-defined minimum is seen in $\rho$.

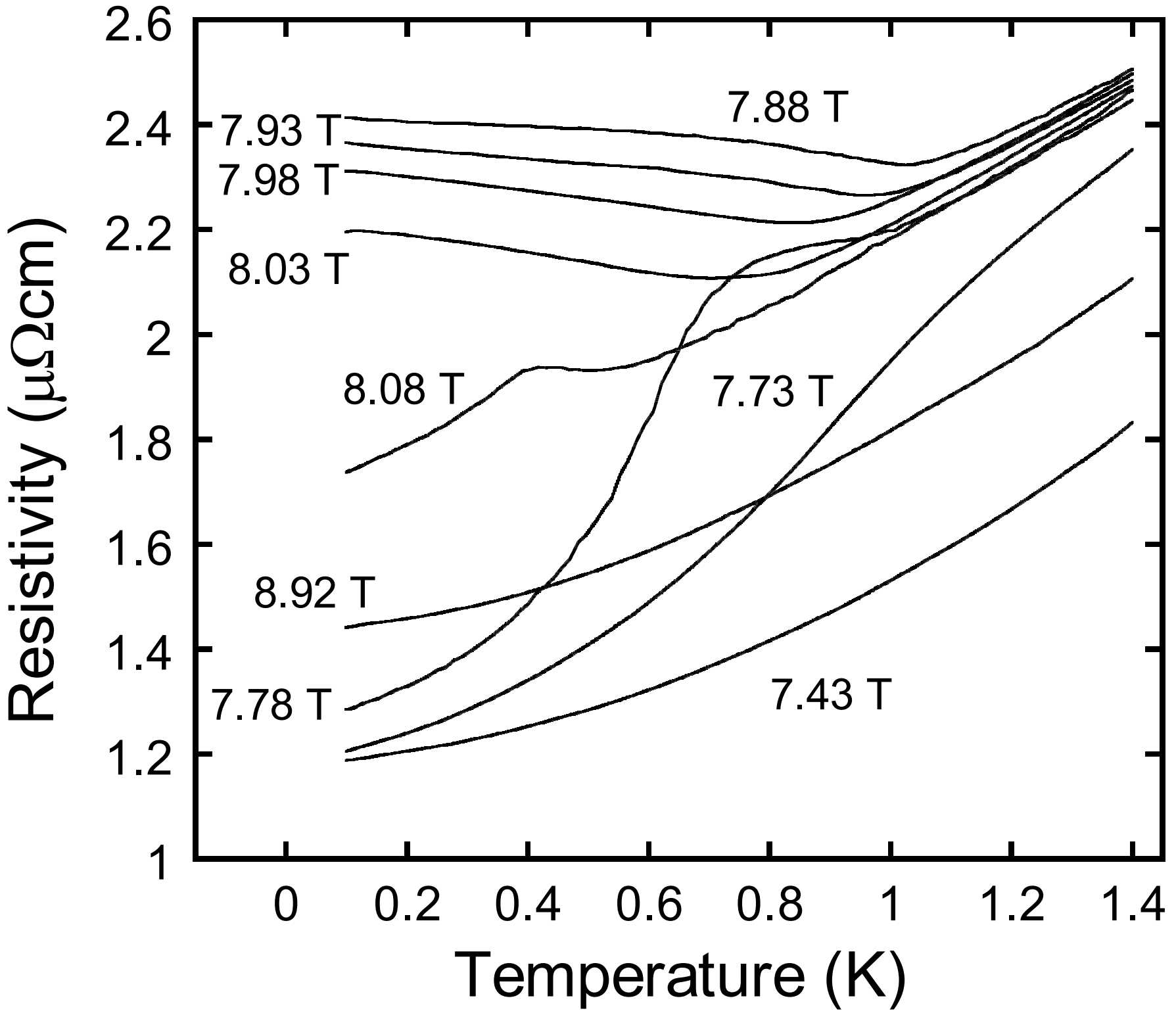

Figure 2. Temperature sweeps of the resistivity at magnetic fields near the metamagnetic transition. The crystal has $\rho_{res}$= 0.55 μΩcm and the sweep rate was approximately 30 mK/min. The measurements were made at Kyoto University, with the samples mounted inside the mixing chamber of a dilution refrigerator to ensure good thermal contact between the samples and the thermometer. For fields below 7.8 T and above 8.1 T, $T^2$ behaviour is seen at low temperatures, but between these fields, an unusual minimum in the resistivity is observed.

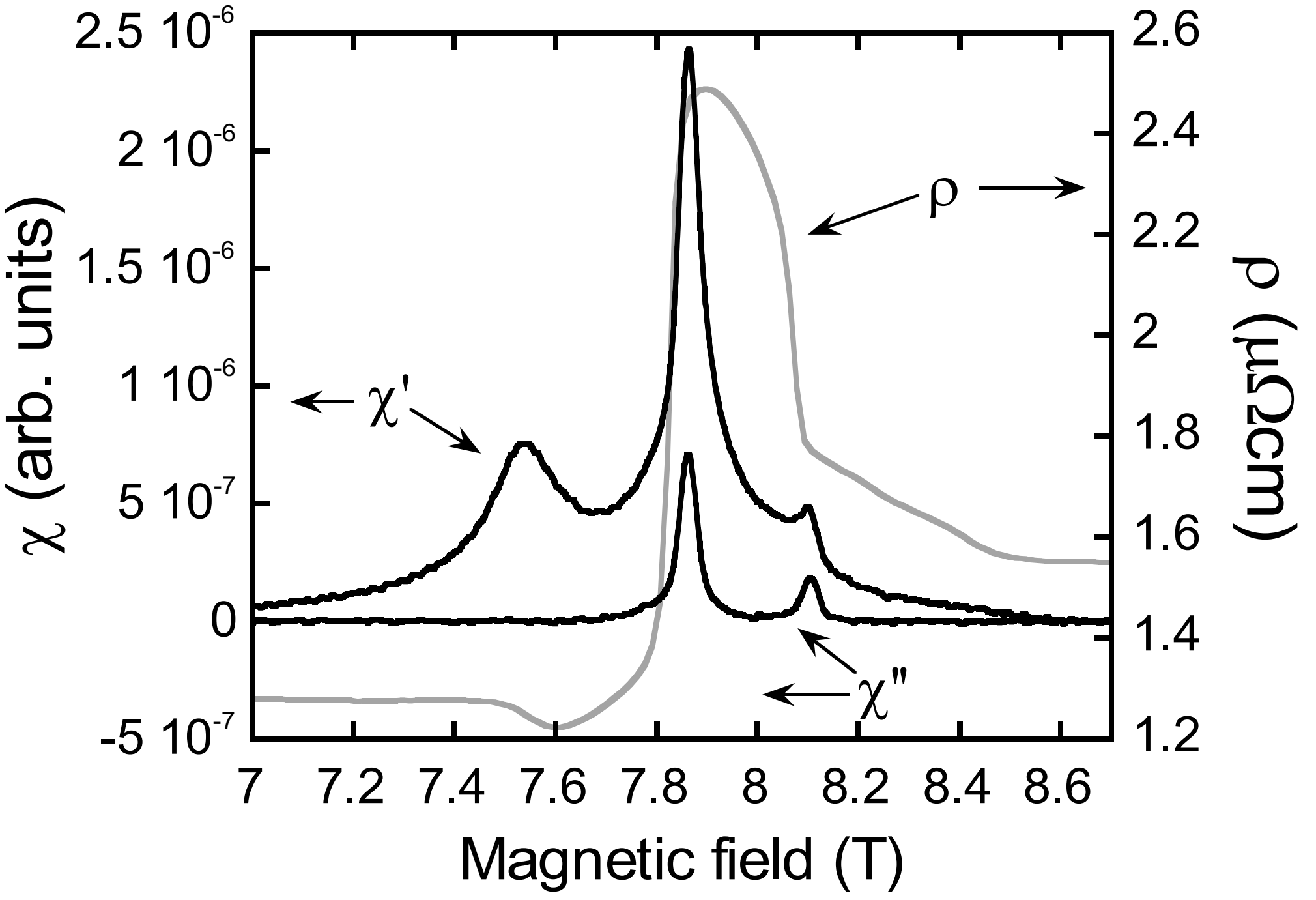

Figure 3. The susceptibility ($\chi$) and resistivity ($\rho$) versus magnetic field across the metamagnetic field. The $\chi$ and $\rho$ measurements were made at $T$ = 20 mK on different crystals, both having $\rho_{res}$< 0.5 μΩcm. The ac excitation field was 0.3 Oe at a frequency of 83.4 Hz. The very sharp changes in $\rho$ are accompanied by peaks in both the real and imaginary parts of $\chi$, indicative of first-order phase boundaries.

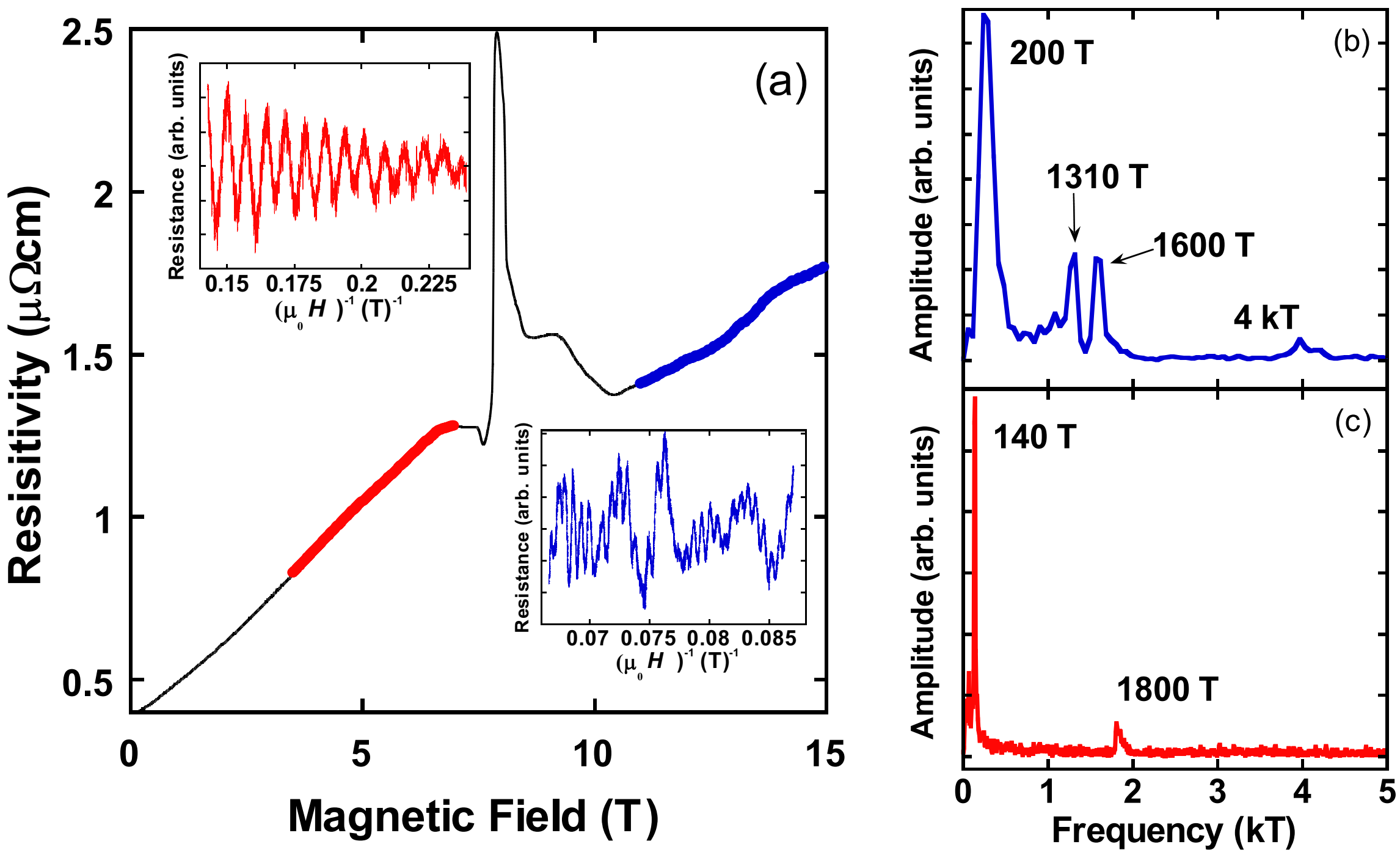

Figure 4. Shubnikov - de Haas oscillations in the resistivity of $Sr_3Ru_2O_7$ in a crystal with $\rho_{res}$ =0.4 μΩcm. (a) The raw data with background subtraction in the insets for the low field (red) and high field (blue) regions, respectively. (b) and (c) show colour coded Fourier transforms of the inset data giving valuable information on the Fermi surface of the metal. The temperature is 0.1 K with the field perpendicular to the $RuO_2$ planes. Similar frequencies were observed in both the 0.4 μΩcm and 0.55 μΩcm crystals.